\documentclass[preprint, showpacs,preprintnumbers,amsmath,amssymb, aip]{revtex4-1}
\usepackage{graphicx}
\usepackage{dcolumn}
\usepackage{bm}
\usepackage{wasysym}
\usepackage{textcomp}
\usepackage{epstopdf}
\usepackage{stmaryrd}
\begin{document}

\preprint{AIP/123-QED}

\title{Optical properties and hardness of highly a-axis oriented AlN films}

\author{Feby Jose}
\affiliation{Thin Film and Coatings Section, Surface and Nanoscience Division, Materials Science Group, Indira Gandhi Centre for Atomic Research, Kalpakkam 603102, India.}

\author{R. Ramaseshan}
\email[]{seshan@igcar.gov.in}
\affiliation{Thin Film and Coatings Section, Surface and Nanoscience Division, Materials Science Group, Indira Gandhi Centre for Atomic Research, Kalpakkam 603102, India.}

\author{S. Tripura Sundari}
\affiliation{Thin Film and Coatings Section, Surface and Nanoscience Division, Materials Science Group, Indira Gandhi Centre for Atomic Research, Kalpakkam 603102, India.}

\author{S Dash}
\affiliation{Thin Film and Coatings Section, Surface and Nanoscience Division, Materials Science Group, Indira Gandhi Centre for Atomic Research, Kalpakkam 603102, India.}

\author{A.K. Tyagi}
\affiliation{Thin Film and Coatings Section, Surface and Nanoscience Division, Materials Science Group, Indira Gandhi Centre for Atomic Research, Kalpakkam 603102, India.}

\author{M.S.R.N. Kiran}
\affiliation{Department of Materials Engineering, Indian Institute of Science, Bangalore, India.}

\author{U. Ramamurty}
\affiliation{Department of Materials Engineering, Indian Institute of Science, Bangalore, India.}

\date{\today}

\begin{abstract}
This paper reports optical and nanomechanical properties of predominantly a-axis oriented AlN thin films. These films were deposited by reactive DC magnetron sputtering technique at an optimal target to substrate distance of 180 mm. X-ray rocking curve (FWHM = 52 arcsec) studies confirmed the preferred orientation. Spectroscopic ellipsometry revealed a refractive index of 1.93 at a wavelength of 546 nm. The hardness and elastic modulus of these films were 17 and 190 GPa, respectively which are much higher than those reported earlier can be useful for piezoelectric films in bulk acoustic wave resonators.

\end{abstract}

\maketitle

Group III nitride semiconductors have attracted a considerable attention in the last two decades due to their unique
 properties such as wide band gap, electrical insulation and high thermal conductivity. These properties have led to potential
 applications in short-wavelength light source/detector and high temperature/frequency devices. \cite{Taniyasu2006, Fu, Handbook, Feby, Wang, Vergara} AlN exists in thermodynamically stable tetrahedrally coordinated wurzite (a=0.3112 nm and c=0.4982 nm) as well as in zinc blende structures.\cite{Fu,Handbook} In recent past, demand for high frequency devices has risen due to the increase in spread of mobile telephones, satellite broadcasting and wireless local area network systems. Dielectric ceramics and surface acoustic wave (SAW) devices have been  conventionally used for fabrication of such high-frequency devices. In addition to the application of oriented AlN in light emitting diode (LED), it has been widely used for the SAW and film bulk acoustic wave (FBAW) devices because of the high surface acoustic wave velocity and suitable piezoelectric coupling factors.\cite{Fu} 

Thin films of AlN  oriented in a-axis (100) and c-axis (002) are important for transverse and longitudinal acoustic wave applications, respectively. They also find application as beneficial buffer layers for the growth of GaN films for optoelectronic devices and hexagonal poly-types of silicon carbide films which have  similar crystal structure and lattice constants.\cite{Deborah,Kanamura}  Majority of research reported to date has been on the synthesis of AlN thin films on various substrates. The ability to grow oriented AlN thin films on Si substrate without a buffer layer can open further opportunities for a variety of application that will exploit the functionality and flexibility of nano-electronic devices. The growth of single domain AlN films on Si (100) is a challenging task for heteroepitaxy due to the mismatch in the lattice parameters and the difference in crystallographic symmetry to the hexagonal close packed AlN (002) lattice plane. Recently, Taniyasu \emph{et al.}\cite{Taniyasu2006} have achieved an AlN PIN (p-type/intrinsic/n-type) homo-junction LED with an emission wavelength of 210 nm using a-plane oriented AlN, which is the shortest reported hitherto. A non c-plane, such as a-plane (100) or M-plane (1010), is expected to enhance the light extraction and exhibit a strong emission intensity along the surface normal in LED structures, which would be superior to that of the (002) oriented plane.\cite{Taniyasu2010} Because of the strong surface orientation of AlN, which influences the emission property, a-plane LEDs are unique and it is important for the improved emission efficiency. Only a few reports are available about the preparation and properties of a-axis oriented AlN films. \cite{Zhi-Xun, X H Ji, Ishihara, Medjani, Hirofumi} The properties of AlN depend  on many factors, such as the technique employed for the deposition, deposition parameters and substrate.\cite{Oliveira, Clement} Reactive magnetron sputtering is an important method to synthesize ceramic thin films at low substrate temperatures, with good  surface finish as well as adhesion.\cite{Feby, Wang, Vergara} There are many optical and mechanical studies on polycrystalline AlN films with c-axis orientation whereas on a-axis oriented AlN films such studies are rather sparse.  \cite{Feby,Wang,Vergara}  In this letter, we  report optical and mechanical properties of a-axis oriented AlN thin films deposited by reactive sputtering. 

AlN thin films of thickness around 350 nm were grown on Si (100) substrate by DC reactive magnetron sputtering equipment (MECA – 2000, France). A 4N pure aluminum target with a diameter of 50 mm in a Ar/N$_2$ gas mixture (4:1) was used. The base pressure of the chamber was less than 6 x $ 10^{-6}$ mbar, while the sputtering pressure was maintained at 5 x $ 10^{-3}$ mbar. The substrates were heated up to $300\,^{\circ}{\rm C}$, whereas the target to substrate distance (TSD) was varied between 100 to 180 mm. X-ray diffraction (XRD) (Bruker D8, Germany) was utilized to evaluate the crystalline quality of AlN thin films. The refractive index and extinction coefficient were derived from the spectroscopic ellipsometry parameters that were measured with a rotating polarizer type instrument (SOPRA ESVG model) in the wavelength range 200 to 900 nm at an incident angle of $75^{\circ}$. The indentation experiments were performed using a nano-indenter (Triboindenter of Hysitron, Minneapolis, USA) with an in-situ imaging capability. A three sided pyramidal diamond Berkovich tip with an end radius of 100 nm was used in this study. The hardness \emph{H} and modulus \emph{E} values reported in this manuscript were evaluated using periodic partial unloading technique. The advantage of this technique is that it estimates the depth dependent variation in the \emph{H} and \emph{E} values from a single indent. In the present case, a peak load \(\ $P$_{max}\) of 1000 $\mu$N was applied to the indenter in 25 loading and unloading steps. The latter were fitted with Oliver-Pharr method to extract \emph{H} and \emph{E} values.\cite{Oliver} Indentation was carried out with loading and unloading rates of 100 $\mu$N/s with a hold of 5 s at \(\ $P$_{max}\).  

Fig.1 (a-d) display the XRD patterns of AlN thin films taken for TSD of 100 nm from RT to $300\,^{\circ}{\rm C}$, while (d-f) show the XRD patterns for TSD 100, 140, 180 nm at $300\,^{\circ}{\rm C}$. Fig.1 (a-e) were recorded in Grazing Incidence (GIXRD) mode. These patterns show poly crystalline hexagonal structure and agree with the JCPDS data 25-1133, corresponding to wurzite AlN. GIXRD mode was also attempted for specimen grown at TSD of 180 nm and $300\,^{\circ}{\rm C}$. However, no pattern could be recorded for this. This led us to speculate that texturing could have occurred and therefore, a B-B geometry was attempted. Fig.1 (f) taken in B-B geometry clearly exhibits the emergence of texturing, showing a strong peak at $33.17^{\circ}  (2\theta)$ corresponding to (100) oriented a-axis AlN. Due to the tensile residual stresses, that start occurring with an increase in TSD, the peak (100) shifts to lower Bragg angle from $33.337^{\circ}  to \ 33.027^{\circ}$.
 
 \begin{figure}[h]
\includegraphics[width=0.45\textwidth]{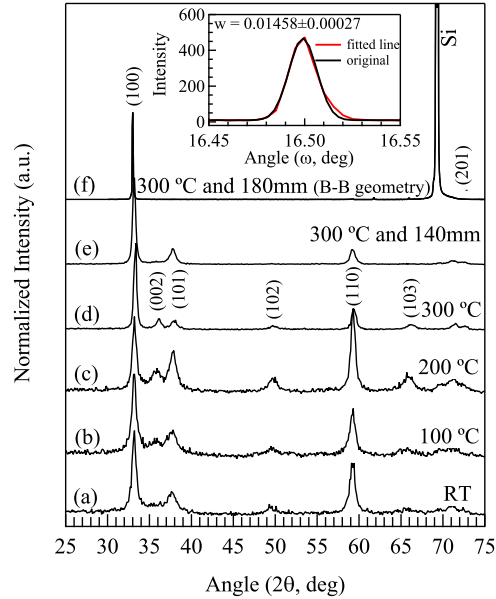}
\caption{\label{Fig1}(Color online) (a-d) display the XRD patterns of AlN thin films taken for TSD of 100 nm from RT to $300\,^{\circ}{\rm C}$, while (d-f) show the XRD patterns for TSD 100, 140 and 180 nm at $300\,^{\circ}{\rm C}$, ((a-e) was recorded in GIXRD mode. Note that (f) from B-B geometry for which the rocking curve profile is shown in the inset with FWHM.}
\end{figure}
 
Fig.1 (inset) shows the X-ray rocking curve of the highly oriented AlN film with TSD 180 mm. XRD $(\omega-2\theta)$ scans confirmed that these films are crystalline and show preferred orientation with wurtzite structure. Full width at half maximum (FWHM) from rocking curve of the AlN (100) reflection at $32.998^{\circ}$ was measured as 52 arcsec. The increase in FWHM of rocking curve is an indication of relative disorientation of the crystals in the film and the lateral crystal size. If the width of the angular range is small, then the diffracted intensity is completely from the perfectly orientated crystals. 

In wurtzite AlN, two types of Al-N bonds exist; namely B1and B2.\cite{Medjani} The energy of formation of B1 type bond is less than the B2 type. (002) and (101) planes are made up of B1 and B2 type bonds whereas (100) plane is constituted only by B1 bonds.  Therefore, adatoms with lower energy content are suitable for the formation of (100) plane.  According to Ishihara \emph{et al.}\cite{Ishihara} the formation of a-axis oriented AlN is enhanced when the TSD is shorter than the mean free path of Al-N dimer. However, this is on the basis of the deposition of a-axis oriented AlN films using the cathodic arc evaporation with a shield, where the energy of the deposited ions is lower.\cite{Hirofumi} The surface and cross sectional image of this highly oriented film is shown in Fig. 2. The surface morphology of this film exhibits a columnar growth with occasional mound like growth on the surface.  The surface of this film with small protrusions that occur at quite uniform distances is indication of the Volmer-Weber island growth mode. 

\begin{figure}[h]
\includegraphics[width=0.45\textwidth]{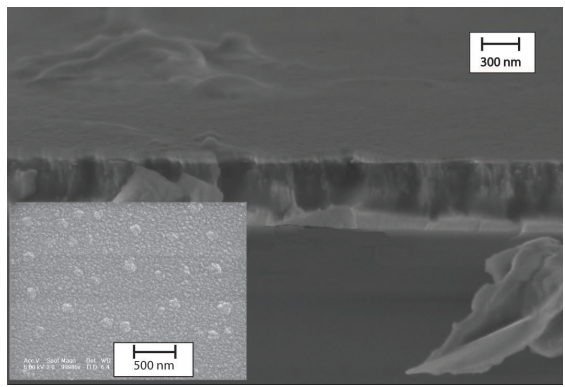}
\caption{\label{Fig2} Cross sectional SEM image of a-axis oriented AlN.}
\end{figure}

The optical response of the film that was synthesized at a substrate temperature of $300\,^{\circ}{\rm C}$ with a TSD of 180 mm which yielded a-axis oriented film (see Fig.1) was examined. Fig. 3 shows its refractive index (\emph{n}) and extinction coefficient (\emph{k}) in the wavelength range of 200 to 900 nm, while the \emph{n} is 1.935 at a wavelength of 546 nm. \emph{k} of the film was less than $10^{-2}$ in this wavelength region, which reveals the transparent nature of the films due to the wide band gap.  Hirofumi \emph{et al.}\cite{Hirofumi} have earlier examined the optical properties of a-axis oriented AlN films. However some ambiguities about their values exist. For example, they  obtained the a-axis oriented film using shielded vacuum arc deposition method at different partial pressures of nitrogen on lattice matched Mo substrate (a=3.147 $\AA$). But their XRD do not indicate to a systematic variation in the intensity of the a-axis peak $(2\theta= 33.2^{\circ})$ with different partial pressure of nitrogen. The \emph{n} and \emph{k} values reported by them are for films coated on quartz substrates. Also the XRD for AlN on Mo substrate alone is shown, however not for AlN on quartz. Therefore, we believe that ours is the first report on a-axis oriented AlN thin films coated on Si (100). A comprehensive optical properties of AlN single crystals as well as  thin films are reported by Loughin et al, for example n is 2.080 at a wavelength of 547 nm.\cite{Loughin}

\begin{figure}[h]
\includegraphics[width=0.45\textwidth]{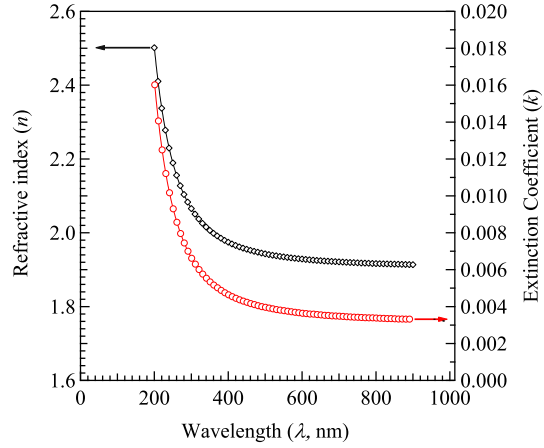}
\caption{\label{Fig3}(Color online) n and k values of a-axis oriented AlN film for the wavelength range of 200 to 900 nm.}
\end{figure}

\begin{figure}[h!]
\includegraphics[width=0.45\textwidth]{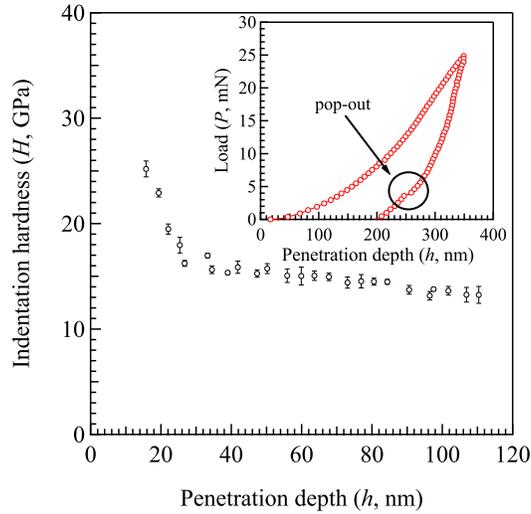}
\caption{\label{Fig4}(Color online) Indentation hardness of a-axis oriented AlN thin film, Inset shows a representative load vs depth penetration curve with the pop-out during unloading highlighted.}
\end{figure}

Design of devices requires, solving the problem of controlling residual stress / strain and the defect  density contained in the thin solid film. The mechanical characterization is vital to get the above mentioned properties to understand the nature of the films. There are reports on evaluation of  elastic stiffness coefficients of AlN and other III-V nitrides by ab-initio calculations. The variation of  \emph{H} with the indenter penetration depth, measured using nanoindentation, is shown in Fig. 4.  As seen \emph{H} values reach a steady after about 30 nm depth of penetration. The initial \emph{H} values (i.e. up to a depth of 30 nm) are not considered because it is known that at shallow penetration depths improper area function can enhance the hardness values. Therefore, we have reported the average \emph{H} and \emph{E} values of the films that have been recorded from 30 - 120 nm depth. The average hardness value in the plateau region is 17 GPa. Similarly the elastic modulus of the films in this region is measured to be 190 GPa. The representative load-displacement curve at higher force (30 mN) is shown an inset in Fig. 4. It shows pop-out behaviour of silicon (film-substrate interface); known to occur due to phase transformation during unloading. However, below \(\ $P$_{max}\) = 1000 $\mu$N no such pop-out in the unloading curve was observed which indicates that the indenter did not penetrate through the underneath Si substrate. Therefore, it is clear that there is no effect of substrate on the mechanical properties of AlN reported in this study. Moreover, the smooth loading nature indicates the absence of crack or heterogeneous deformation during application of load. In summary, we have observed the TSD and substrate temperature influence the orientation of the AlN films. The quality of grain orientation was analyzed using XRD rocking curve. The refractive index of a-axis oriented AlN film by ellipsometry is 1.95 for a wavelength of 546 nm.

 F.J would like to acknowledge Dr. Ramanathasamy Pandian for the SEM images. M.S.R.N.K. thanks University Grants Commission, Govt of India, for Dr. D. S. Kothari Post-Doctoral Research Fellowship.

\end{document}